# Transport studies of electron-hole and spin-orbit interaction in GaSb/InAsSb core-shell nanowire quantum dots

*Bahram Ganjipour[1,*], Martin Leijnse[1], Lars Samuelson[1], H.Q. Xu[1,2], Claes Thelander[1]*

[1] Division of Solid State Physics, Lund University, Box. 118, S-22100, Lund, Sweden

[2] Department of Electronics and Key Laboratory for the Physics and Chemistry of Nanodevices, Peking University, Beijing, China

Corresponding author: bahram.ganjipour@gmail.com

## ABSTRACT

We report low-temperature transport studies of parallel double quantum dots formed in GaSb/InAsSb core-shell nanowires. At negative gate voltages, regular patterns of Coulomb diamonds are observed in the charge stability diagrams, which we ascribe to single-hole tunneling through a quantum dot in the GaSb core. As the gate voltage increases, the measured charge stability diagram indicates the appearance of an additional quantum dot, which we suggest is an electron quantum dot formed in the InAsSb shell. We find that an electron-hole interaction induces shifts of transport resonances in the source-drain voltage from which an average electron-hole interaction strength of 2.9±0.3 meV is extracted. We also carry out magnetotransport measurements of a hole quantum dot in the GaSb core and extract level-dependent *g*- factors and a spin-orbit interaction.

The GaSb/InAs heterostructure has been investigated for various device implementations since the first pioneering work by Sakaki et al. in the late 1970's.[1, 2, 3] Recently, the heterostructure has gained an increased interest in fundamental quantum physics studies [4, 5, 6] because of its exotic type-II broken band alignment, in which the bottom of the InAs conduction band lies around 150 meV below the top of the GaSb valence band.[7] This leads to the coexistence of spatially separated electrons (in InAs) and holes (in GaSb)[8] and also to electron-hole hybridization across the heterointerface.[9] For GaSb/InAs quantum well layers it is possible to tune the quantum confinements of electrons and holes in the growth direction, such that only the first electron subband in the InAs overlaps with the highest hole subband in the GaSb[10, 11] and, consequently, a small hybridization gap opens up in the band structure.[12, 13] It has been proposed that such band-inverted, hybridized InAs/GaSb quantum wells can host quantum spin Hall states.[6] Such states have been probed in experiments that show that charge transport in the interface is dominantly carried by edge channels.[11, 14] Other studies have investigated the existence of helical edge states inside the hybridization gap using InAs/GaSb four-terminal devices with various dimensions.[4, 15] It has also been shown that GaSb/InAs quantum wells form a good interface to superconductors.[16] Interest in this material system has furthermore been stimulated by the formation of indirect excitons[17, 18, 19] which can be used for many optoelectronics applications[20] and for studies of exciton condensations[21] because of their long life-time.

Here we focus on a much less investigated form of this material system—GaSb nanowires surrounded by a thin As-rich InAsSb shell. The unique electronic properties of the GaSb/InAs semiconductor heterostructure, combined with the one-dimensional (1D) geometry of nanowires, provides a promising platform for future exciton- and spin-physics studies, including the search for an excitonic superfluidity.[21] Recently, we investigated the electronic properties of GaSb/InAsSb core-shell nanowires with a range of shell thicknesses. We found that nanowires with a core diameter of 50-70 nm and an InAsSb shell thickness of 5-7 nm exhibit ambipolar conduction behavior with a finite conductance valley at low temperatures[22], implying a broken band gap alignment in the wires[23] which is a prerequisite for the hybridization of electron and hole states.[24] It was also shown that nanowires with an even thinner InAsSb shell exhibit dominant $p$-type semiconducting characteristics, with improved hole transport properties when compared to bare GaSb nanowires.[25] These nanowire structures can also be used to study magnetotransport properties of confined holes. With the intrinsic strong spin-orbit interaction of holes in the valence band, these 1D semiconductor nanowires can provide an attractive platform for the study of Majorana bound states in the solid state when combined with an $s$-wave superconductor.[26]

In this Rapid communication we observe evidence of spatially separated parallel hole and electron quantum dots formed within the valence band of GaSb and the conduction bands of InAsSb, respectively. We find a strong Coulomb interaction between electrons and holes, which is promising, for example, for Coulomb drag experiments and for the realization of excitonic condensation which has recently been proposed theoretically for the GaSb/InAs heterojunction both in the core-shell nanowire geometry [21] and in the geometry of quantum wells.[18] In the hole transport regime of such core-shell quantum dots, pronounced excited states are observed, where the magnetic field dependence of the differential conductance peaks reveals a two-fold spin degeneracy, as expected for strongly confined holes. Zeeman splittings of differential conductance peaks are used to extract $g$-factors and we also observe an avoided level crossing, from which a spin-orbit interaction strength of 75 $\mu$eV is deduced.

The heterostructure GaSb/InAsSb nanowires are grown using metalorganic vapor phase epitaxy from Au aerosol nanoparticles deposited on GaAs (111)B substrates. [27, 28] Figures 1(a) and 1(b) show low- and high-resolution transmission electron microscopy (TEM) images of a grown GaSb/InAsSb core-shell nanowire. It is seen in Fig. 1(b) that an InAsSb shell is formed around the GaSb core during growth of the axial InAs segment. The shell composition is thus not pure InAs, but an As-rich ternary InAs$_{0.8-0.9}$Sb$_{0.2-0.1}$, due to a memory effect of Sb in the reactor.[27] However, for simplicity, from hereon, we refer to this ternary as InAs in the text. The nanowires have a GaSb core diameter of 50-70 nm, overgrown with an InAs shell of around $4 \pm 1$ nm in thickness.

The experimental results reported here are obtained from two separate samples[29]—the first characterized in a $^3$He refrigerator with a base temperature of 300 mK (device $A$) and the second in a dilution refrigerator with a base temperature of 50 mK (device $B$).

Figure 1(c) shows a scanning electron microscope (SEM) image of device $A$ fabricated from a GaSb/InAs core-shell nanowire with a diameter of ~60 nm. Room temperature back-gate transfer characteristics show typical hole transport behavior for gate voltages smaller than 6 V [30] resulting in the formation of a hole quantum dot at low temperature for gate voltages smaller than 6 V, likely within the valence band of the GaSb core. The differential conductance, $dI_{sd}/dV_{sd}$, as a function of $V_{bg}$ and $V_{sd}$ (charge stability diagram) is measured for device $A$ and plotted on a logarithmic scale in Fig. 2(a), showing a set of regular Coulomb diamond structures. All diamond edges have the same slopes and all diamonds close at $V_{sd} = 0$ V, indicating the formation of a single dot between the contacts. The lines of high differential conductance running parallel to the diamond edges (marked with dashed white lines) in Fig. 2(a) are related to sequential tunneling through excited states of the dot. The device shows a set of Coulomb diamonds with irregular shapes at higher gate voltages [Fig. 2(b)], where the slopes of the edges of the adjacent diamonds are not parallel, in strong contrast to the regular parallelograms observed in Fig. 2(a). As we will argue below, this unconventional charge stability diagram is a sign of the formation of a parallel double dot consisting of an electron quantum dot located in the InAs shell and a hole quantum located in the GaSb core.

In order to study the excited state spectrum and explore the origin of such unusual Coulomb diamonds, device $B$ was fabricated based on a similar GaSb/InAs core-shell nanowire and cooled down in a dilution refrigerator with a base temperature of 50 mK. The device is first investigated in the hole regime (less positive gate voltage) [31] where we observe mostly regular Coulomb diamonds, see Fig. 3(a).

For more positive gate voltage region, with fewer holes in the quantum dot, we observe more of the irregular coulomb diamonds that appear in both devices. Here for device $B$ as shown in Figure 3(b), we see two sets of diamonds with different edge slopes super-imposed on each other, indicating the formation of a double quantum dot, where the two dots have different capacitive couplings to the source, drain, and gate. In long nanowire devices we often observe the unintentional formation of a double quantum dot with the two dots coupled in series, for which zero-bias tunneling is only possible when both dots happen to be at resonance. In the present case all diamonds close at $V_{sd}= 0$ V, evidencing instead a double dot with the two dots coupled in parallel with respect to the source and drain because tunneling is possible even if only one quantum dot is at resonance. In this material system the most likely explanation for this observation is that an electron quantum dot forms in the InAs shell and a hole quantum dot forms in the GaSb core, which can coexist for a certain gate voltage span. A similar diamond pattern has been observed in Si-based hole quantum dots, which was attributed to the coexistence of light-hole and heavy-hole states.[32] This explanation is unlikely in our experiment because the second diamond pattern appears for more positive gate voltages, which is consistent with the interpretation of the formation of an additional electron dot, but opposite to the expectation for the coexistence of light holes and heavy holes.[32] Assuming parallel electron and hole quantum dots, the different slopes of the two diamond patterns are easily understood because the capacitive couplings to all electrodes are likely to be different for the core and shell. The diamonds with lower gate coupling (smaller slope of the edges as a function of $V_{bg}$) are found also for smaller $V_{bg}$ and we interpret these as originating from the hole quantum dot, while the diamond pattern with a larger gate coupling appears only for larger $V_{bg}$ and is interpreted as originating from the electron quantum dot.

Upon closer inspection, it can be seen that the two Coulomb diamond patterns are not completely independent. Figure 3(c) zooms in on the part of Fig. 3(b) indicated by the white dashed rectangle. Going across the leftmost zero-bias crossing point from smaller to larger $V_{bg}$ corresponds to removing a hole from the core, going from a hole population of $n_h$ to $n_h$-1 on the dot. The finite bias lines, marked by black dashed lines extending from this crossing point, correspond to the possibility of a fluctuation in the hole population between $n_h$ and $n_h$-1. Going across one of the two following zero-bias crossing points corresponds to adding one electron to the shell. It can be seen that each time an electron is added, the above mentioned resonance line for the $n_h \Leftrightarrow n_h$-1 fluctuation jumps to a lower bias voltage (lower energy), giving rise to multiple copies of that resonance (all marked with black dashed lines). This jump of the hole resonance line in the bias voltage provides a direct evidence of an attractive interaction between the spatially separated electrons and holes: Every time an electron is added to the shell, the energy needed to add a hole to the core is reduced. An average electron-hole interaction strength of $\Delta_{e\text{-}h}$= 2.9±0.3 meV is obtained from the jumps in hole resonance lines, e.g., the vertical solid black line in Fig. 3(c). The opposite patterns occur for larger $V_{bg}$ in the rightmost part of Fig. 3(b), where resonances corresponding to removing a hole, marked with white dashed lines, jumps to higher energies every time an electron is added. Parallel electron and hole quantum dots can also explain the data from device $A$, where we have also observed a transition to irregular Coulomb diamonds with multiple slopes at large $V_{bg}$ [Fig. 2(b)]. Data from an additional sample is shown in the Supplemental Material, which seem to exhibits the same physics, but where our interpretation is that the hole quantum dot has an even smaller gate coupling and the shell seems to be only very weakly tunnel coupled to the source and drain, making the electron-related Coulomb diamonds rather invisible (Fig. S7).

To further support the above interpretation we have carried out master equation calculations on a double quantum dot system with an electron dot and a hole dot in parallel. The result presented in Fig. 3(d) matches all the essential features of the experimental data in Fig. 3(c), in particular the jumps in the hole resonance lines due to the electron-hole interaction, also marked here with dashed black lines. The model only includes a single spin-degenerate level for an electron and a single spin-degenerate level for a hole, and therefore can only reproduce the diamond edges, not the additional features originating from excited states. A simulated charge stability diagram of a non-interacting parallel quantum dot is shown in the Supplemental Material, Fig. S10(c).[33]

An additional interesting detail is that the electron and hole resonances simply cross each other as seen in the regions indicated by two white rectangles in Fig. 3(c). As was observed in Refs. 34 and 35, a tunnel coupling between the two dots in a parallel double quantum dot gives rise to anti-crossings of the resonance lines. The level crossings seen in our data therefore indicate that there is no visible electron-hole hybridization in this particular type of core-shell double quantum dot. This could be related to the rather high positive gate voltages required to populate the electron quantum dot, which polarizes the two-dot system, and leads to a reduced spatial overlap between the wave functions of the two dots.

In order to extract more information concerning the nature of the quantum dot states, we also studied the magnetotransport properties of device $B$ under a perpendicular applied magnetic field in the hole transport regime (less positive gate voltages). Figure 4(a) shows a charge stability diagram of device $B$ on a linear scale. The magnetic field evolution of the $dI_{sd}/dV_{sd}$ peaks at a fixed gate voltage of 5.019 V, i.e., along cut A, are shown in Fig. 4(b). The application of a magnetic field reveals a two-fold spin-degeneracy of confined hole states as expected, and each resonance splits into a spin-up state and a spin-down state, which is consistent with other experiments on hole quantum dots.[36, 37] At low magnetic field, the Zeeman splitting increases linearly with the field. However, at intermediate magnetic fields (around 2.1 T) the spin-down ground state and spin-up excited state (although the splitting of the excited state is not resolved) undergo an avoided level-crossing as a result of spin-orbit interaction. An energy gap of $\Delta E= 150$ μeV is extracted from the anti-crossing from which the spin-orbit energy $\Delta_{SO} = 75$ μeV is tentatively extracted. This value is smaller than that of their spin-1/2 electronic counterpart as expected for strongly confined hole states.[38, 39]

Zeeman splittings of differential conductance peaks in Figs. 4(b),- 4(d) are used to estimate the hole $g$-factors associated with the corresponding quantum levels. Using the measured Zeeman splittings in the low magnetic field region, we find |$g$*|=1.17±0.3, 3±0.1 and 4.7±0.1.[40] The measurements show that the $g$-factors are quantum level dependent, which we connect to the presence of spin-orbit interaction.[41] Here, the hole $g$-factors in the GaSb core may be affected by the InAs shell because the tail of the wave function may penetrate into the InAs shell which has a bulk $g$-factor of |$g$|=14.[42]

In conclusion, we have studied charge transport in GaSb/InAs nanowire core-shell quantum dots, where the coexistence of spatially separated electron and hole quantum dots with a strong Coulomb interaction between the electron and hole states is observed. An average value of $\Delta_{e-h}= 2.9\pm0.3$ meV is deduced for the electron-hole Coulomb interaction. Master equation calculations for a parallel, core-shell double quantum dot have reproduced the observed Coulomb diamond patterns and the electron-hole interaction characteristics seen in the experiments. From magneto-transport measurements in the regime dominated by holes, an anti-crossing between quantum levels is observed. We attribute this, and the observed level-dependent $g$-factors, to the presence of spin-orbit interaction. We expect that our results will stimulate further works toward the utilization of this exotic materials system for studying electron-hole hybridization, 1D excitonic superfluidity, and Coulomb drag phenomena.

ACKNOWLEDGMENT

We thank L.-E. Wernersson, B. M. Borg, and K. A. Dick for providing the nanowire material and M. Ek for TEM imaging. This work was conducted within the Nanometer Structure Consortium at Lund University (nmC@LU), with financial support from the Swedish Research Council (VR), the Swedish Foundation for Strategic Research (SSF), and the Knut and Alice Wallenberg Foundation (KAW). H.Q.Xu acknowledges also financial supports from the National Basic Research Program of China (Grants No. 2012CB932703 and No. 2012CB932700) and the National Natural Science Foundation of China (Grants No. 91221202, No. 91421303, and No. 61321001).

ASSOCIATED CONTENT

Simulated charge stability diagram of a non-interacting parallel double quantum dot, room temperature and low temperature $I$-$V_{bg}$ characteristics of device $A$. Low temperature $I$-$V_{bg}$ characteristics and charge stability diagram of device $B$ and additional samples.

AUTHOR INFORMATION

REFERENCES

(1) H. Sakaki, L. L. Chang, R. Ludeke, C. A. Chang, G. A. Sai-Halasz, L. Esaki, Appl. Phys. Lett. 31, 211 (1977).

(2) M. Razeghi, S. Abdollahipour, E. K. Huang, G. Chen, A. Haddadi, B. M. Nguyen, Opto-electronics review 19, 46 (2011).

(3) G. Zhou, R. Li, T. Vasen, M. Qi, S. Chae, Y. Lu, Q. Zhang, H. Zhu, J.-M. Kuo, T. Kosel, M. Wistey, P. Fay, A. Seabaugh, H. Xing, International Electron Devices Meeting (2012), San Francisco, CA. Pages 32.61-32.6.4.

(4) I. Knez, R.- R. Du, G. Sullivan, Phys. Rev. B 81, 201301 (2010).

(5) F. Nichele, A. Nath Pal, P. Pietsch, T. Ihn, K. Ensslin, C. Charpentier, W. Wegscheider, Phys. Rev. Lett. 112, 036802 (2014).

(6) C. Liu, T. L. Hughes, X.-L. Qi, K. Wang, S.-C. Zhang, Phys. Rev. Lett. 100, 236601 (2008).

(7) H. Kroemer, Physica E: Low-dimensional Systems and Nanostructures 20, 196 (2004).

(8) M. Altarelli, Phys. Rev. B  28, 842 (1983).

(9) A. Zakharova, S. T. Yen, K. A. Chao, Phys. Rev. B 64, 235332 (2001).

(10) G. H. Sai-Halasz, E. Esaki, W. A. Harrison, Phys. Rev. B 18, 2812 (1978).

(11) E. M. Spanton, K. C. Nowack, L. Du, G. Sullivan, R.-R. Du, K. A. Moler, Phys. Rev. Lett. 113, 026804 (2014).

(12) M. Lakrimi, S. Khym, R. J. Nicholas, D. M. Symons, F. M. Peeters, N. J. Mason, P. J. Walker, Phys. Rev. Lett. 79, 3034 (1997).

(13) M. J. Yang, C. H. Yang, B. R. Bennett, B. V. Shanabrook, Phys. Rev. Lett. 78, 4613 (1997).

(14) I. Knez, R.- R. Du, G. Sullivan, Phys. Rev. Lett. 79, 136603 (2011).

(15) I. Knez, C. T. Rettner, S.-H. Yang, S. S. P. Parkin, L. Du, R.-R. Du, G. Sullivan, Phys. Rev. Lett. 112, 026602 (2014).

(16) I. Knez, R.-R. Du, G. Sullivan, Phys. Rev. Lett. 109, 186603 (2012).

(17) Y. Naveh, B. Laikhtman, Phys. Rev. Lett. 77, 900 (1996).

(18) D. P. Pikulin, T. Hyart, Phys. Rev. Lett. 112, 176403 (2014).

(19) J.-P. Cheng, J. Kono, B. D. McCombe, I. Lo, W. C. Mitchel, C. E. Stutz, Phys. Rev. Lett. 74, 450 (1995).

(20) S. De-Leon, B. Laikhtman, Phys. Rev. B 61, 2874 (2000).

(21) D. S. L. Abergel, arXiv:1408.7065v1, (2014).

(22) B. Ganjipour, M. Ek, B. M. Borg, K. A. Dick, M.-E. Pistol, L.-E. Wernersson, C. Thelander, Appl. Phys. Lett. 101, 103501 (2012).

(23) See Fig. S1 in Supplemental Material at  for the radial band structure.

(24)V. V. R. Kishore, B. Partoens, F. M. Peeters, Phys. Rev. B 86, 165439 (2012).

(25) B. Ganjipour, S. Sepehri, A. W. Dey, O. Tizno, B. M. Borg, K. A. Dick, L. Samuelson, L.-E. Wernersson, C. Thelander, Nanotechnology 25, 425201 (2014).

(26) C. Kloeffel, M. Trif, D. Loss, Phys. Rev. B 84, 195314 (2011).

(27) B. Ganjipour, A. W. Dey, B. M. Borg, M. Ek, M.-E. Pistol, K. A Dick, L.-E. Wernersson, C. Thelander, Nano Lett. 11, 4222 (2011).

(28) B. M. Borg, M. Ek, B. Ganjipour, A. W. Dey, K. A. Dick, L.-E. Wernersson, C. Thelander, Appl. Phys. Lett. 101, 04508 (2012).

(29) See Supplemental Material at    for details on device fabrication.

(30)    See Fig. S2 in Supplemental Material at for room temperature transfer characteristic of a fabricated device from the same growth.

(31) See Fig. S3 in Supplemental Material at   for further information.

(32) F. A. Zwanenburg, C. E. W. M. Van Rijmenam, Y. Fang, C. M. Lieber, L. P. Kouwenhoven, Nano Lett. 9, 1071 (2009).

(33) See Supplemental Material at  for further details on the model.

(34) K. Goss, S. Smerat, M. Leijnse, M. R. Wegewijs, C. M. Schneider, C.  Meyer, Phys. Rev. B 83, 201403 (2011).

(35) A. Eliasen, J. Paaske, K. Flensberg, S. Smerat, M.; Leijnse, M. R. Wegewijs, H. I. Jørgensen, M. Monthioux, J. Nygård, Phys. Rev. B 81, 155431 (2010).

(36) B. D. Gerardot, D. Brunner, P. A. Dalgarno, P. Öhberg, S. Seidl, M. Kroner, K. Karrai, N. G. Stoltz, P. M. Petroff, R. J. Warburton, Nature 451, 441 (2008).

(37) H. Hu, F. Kuemmeth, M. C. Lieber, M. C. Marcus, Nature Nanotechnology 7, 47 (2012).

(38) D. V. Bulaev, D. Loss, Phys. Rev. Lett. 95, 076805 (2005).

(39) D. Heiss, S. Schaeck, H. Huebl, M. Bichler, G. Abstreiter, J. J. Finley, D. V. Bulaev, D. Loss, Phys. Rev. B 76, 241306 (2007).

(40) See Fig. S9 in Supplemental Material at   for more information on the extracted $g$-factors.

(41) H. A. Nilsson, P. Caroff, C. Thelander, M. Larsson, J. B. Wagner, L.-E. Wernersson, L. Samuelson, and H. Q. Xu, Nano Lett. 9, 3151 (2009).

(42) S. Csonka, L. Hofstetter, F. Freitag, S. Oberholzer, C. Schönenberger, T. S. Jespersen, M. Aagesen, J. Nygård, Nano Lett. 8, 3932 (2008).

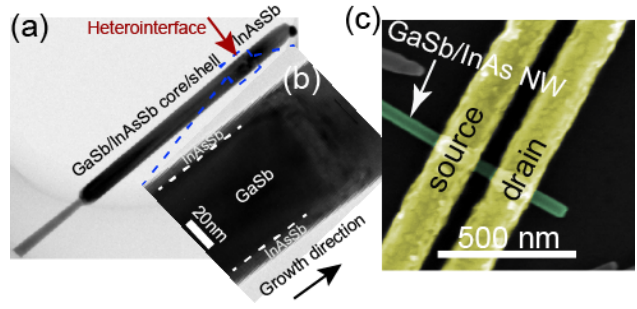

FIG. 1: (a) Low- and (b) high-resolution TEM images of a core-shell nanowire showing the presence of a 6-nm-thick As-rich InAsSb shell around the GaSb core. The white dashed lines indicate the GaSb/InAsSb radial heterointerface. (c) SEM image of a core-shell nanowire quantum dot device (device *A*).

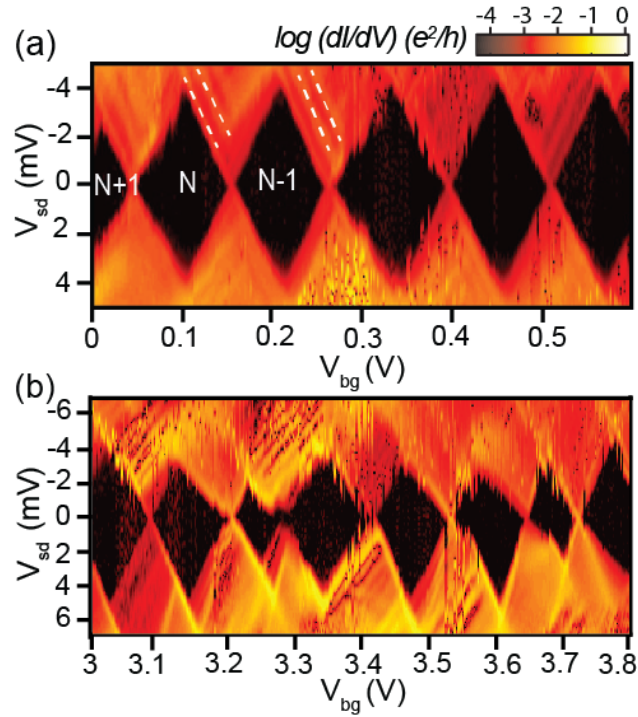

FIG. 2: (a) Charge stability diagram (logarithmic color scale) as a function of $V_{sd}$ and $V_{bg}$ deep in the hole transport regime, where a conventional Coulomb diamond pattern is observed. An addition energy of 4 meV is extracted from the size of the diamonds, corresponding to a total dot capacitance of $C_{\Sigma} = 40$ aF. The back-gate lever arm and capacitance are found to be $\alpha_{bg} = 0.035$ and $C_{bg} = 1.39$ aF, respectively. (b) Charge stability diagram (logarithmic color scale) of the device for larger positive gate voltages showing a set of Coulomb diamonds where the slopes of the edges in adjacent diamonds are not parallel.

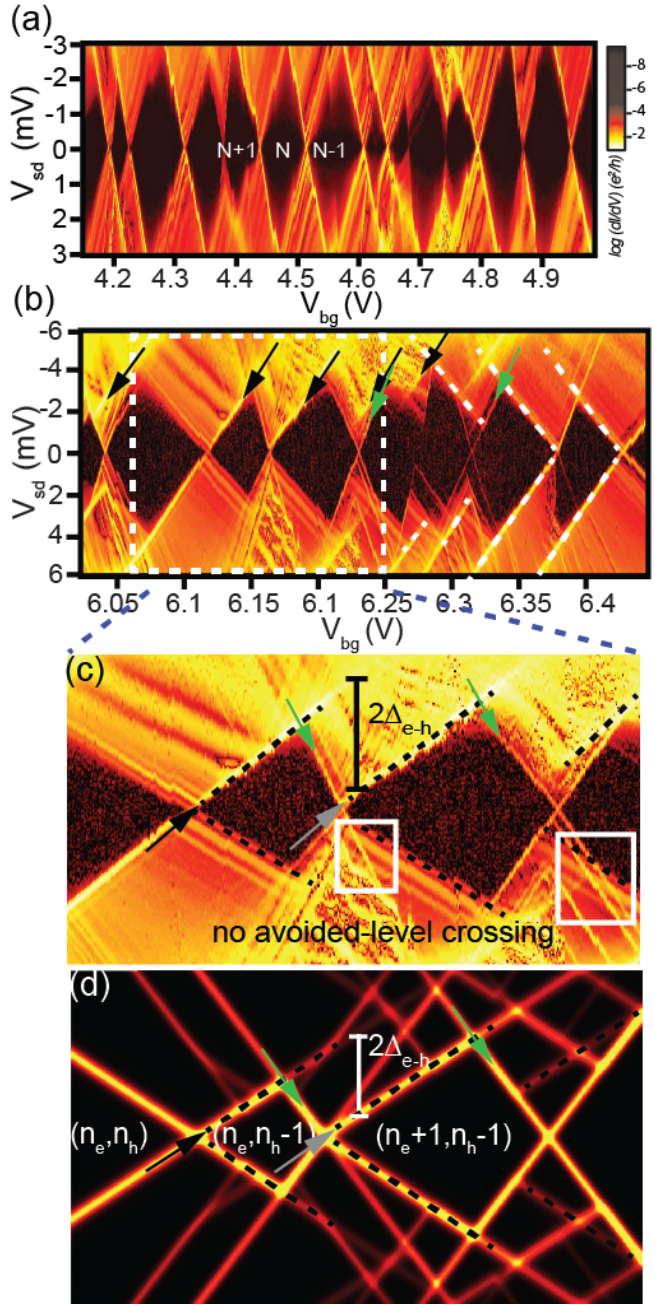

FIG. 3: (a) Charge stability diagram (logaritmic color scale) of device *B* in the hole regime where we observe mostly regular Coulomb diamonds. (b) Charge stability diagram (logarithmic color scale) of device *B* for large gate voltages, showing two patterns of diamonds with different slopes (indicated by black and green arrows) superimposed on each other. (c) Higher resolution scan of the area indicated with a white dashed rectangle in panel (b). An average electron-hole interaction $\Delta_{e\text{-}h}$= 2.9±0.3 meV is extracted from the jumps in the hole resonance as an electron is added. The hole and electron states

show no anti-crossing energy gap as indicated by the white rectangles, implying that the quantum dots are not tunnel coupled. (d) Simulated charge stability diagram of a pair of Coulomb-interacting parallel hole and electron quantum dots with occupation numbers shown inside the corresponding diamonds. The gray arrow represents the same hole state marked by the black arrow when one electron is added to the system.

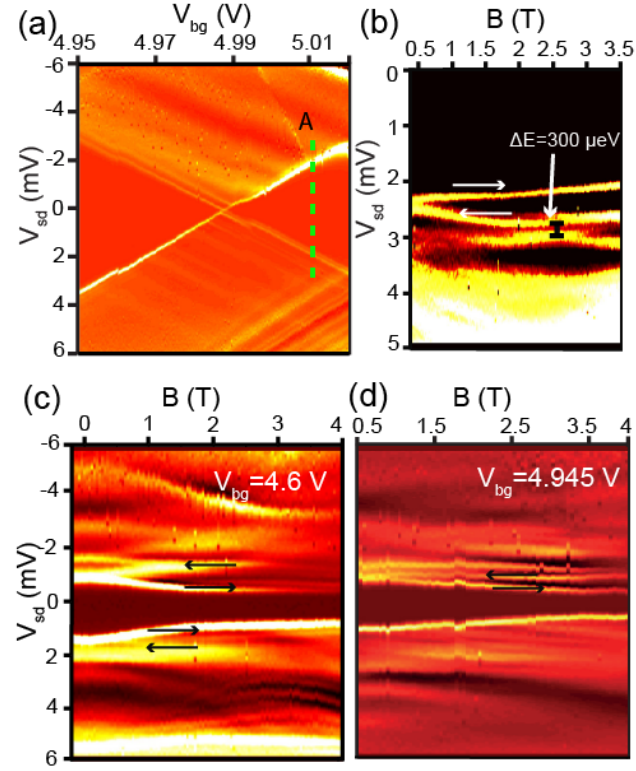

FIG. 4: (a) Charge stability diagram of device $B$ on a linear scale. (b) $dI_{sd}/dV_{sd}$ as a function of $V_{sd}$ and $B$ at $V_{bg} = 5.019$ V, i.e., along cut A in panel (a), showing an avoided level crossing. An anti-crossing energy gap of $\Delta E_{anti\text{-}cross} = \Delta E/2 = 150$ µeV is extracted corresponding to $\Delta_{SO} = 75$ µeV. Due to symmetrical biasing, the anti-crossing gap is half of the measured energy gap in panel (b). A linear fit to the data yields a value of |$g*$|= 3±0.1 associated with this quantum level. (c) and (d) $dI_{sd}/dV_{sd}$ as a function of $V_{sd}$ and $B$ at $V_{bg}$= 4.6 and 4.945 V. Linear fits to the Zeeman splittings (marked with arrows in the upper part of the panels) yields $g$-factors of  |$g*$|= 4.7±0.1 and |$g*$|= 1.17±0.3.

# Transport studies of electron-hole and spin-orbit interaction in GaSb/InAsSb core-shell nanowire quantum dots

*Bahram Ganjipour[1,*], Martin Leijnse[1], Lars Samuelson[1], Hongqi Xu[1,2], Claes Thelander[1]*

[1] Division of Solid State Physics, Lund University, Box. 118, S-22100, Lund, Sweden

[2] Department of Electronics and Key Laboratory for the Physics and Chemistry of Nanodevices, Peking University, Beijing, China

*Corresponding author's e-mail. bahram.ganjipour@gmail.com

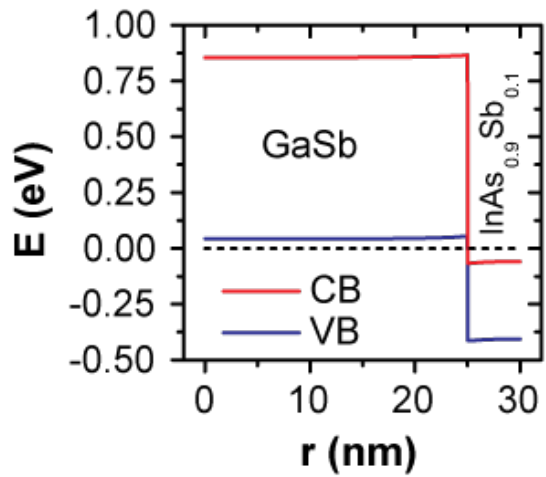

Fig. S1. Poisson simulation (Ref. 27) of the radial band structure ($T$ = 4K), without confinement, for a GaSb-InAs$_{0.9}$Sb$_{0.1}$ heterostructure, with a 50 nm diameter core, and a 5 nm thick shell.

**Device Fabrication:** Nanowires are transferred from the growth substrate to a degenerately doped Si substrate with a 100 nm SiO$_2$ cap layer. Two Ti/Au (20/80 nm) contacts with varying spacings are defined on the GaSb/InAs core-shell nanowire segments using electron beam lithography, followed by a buffered HF etch before metallization.

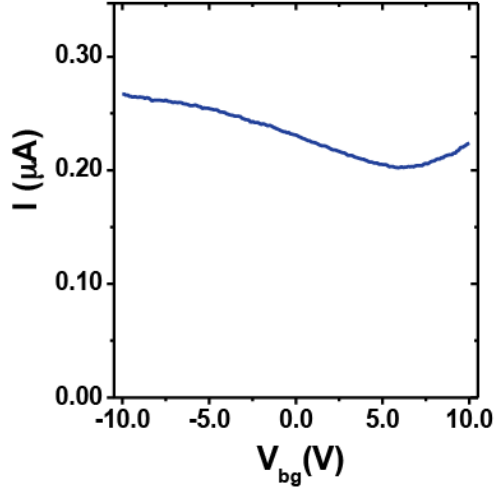

Figure S2: Room temperature back-gate transfer characteristics of a GaSb/InAs core/shell nanowire from the same growth batch as device *A* for source–drain bias voltage of $V_{sd} = 10$ mV showing typical hole transport behavior for gate voltage smaller than 6 V.

**Transport measurements:** Source-drain current of device *B* as a function of back-gate voltage measured with $V_{sd} = 100$ $\mu$V at the base temperature of 50 mK. The quantum dot exhibits Coulomb blockade oscillations, where the Coulomb peak spacing decreases with decreasing gate voltage, which provides a strong evidence that a hole quantum dot is formed between the source and drain contacts.

Fig. S3(c) shows a high-resolution scan of one Coulomb diamond (diamond $N$th of the Fig. 3(a) in the main paper) with pronounced excited state conductance lines. A back-gate capacitance of $C_{bg}= 2.1$ aF, a gate lever arm of $\alpha_{bg} = 0.045$ and a total dot capacitance of $C_\Sigma = 48$ aF are obtained from the diamond labeled $N$. The high differential conductance lines running parallel to the edges of the diamond are results of transport through excited states. Single-particle energy-level spacings ($\Delta$E), with an average value of 0.26 meV, are extracted from Fig. S3(c) and are plotted in Fig. S3(d).

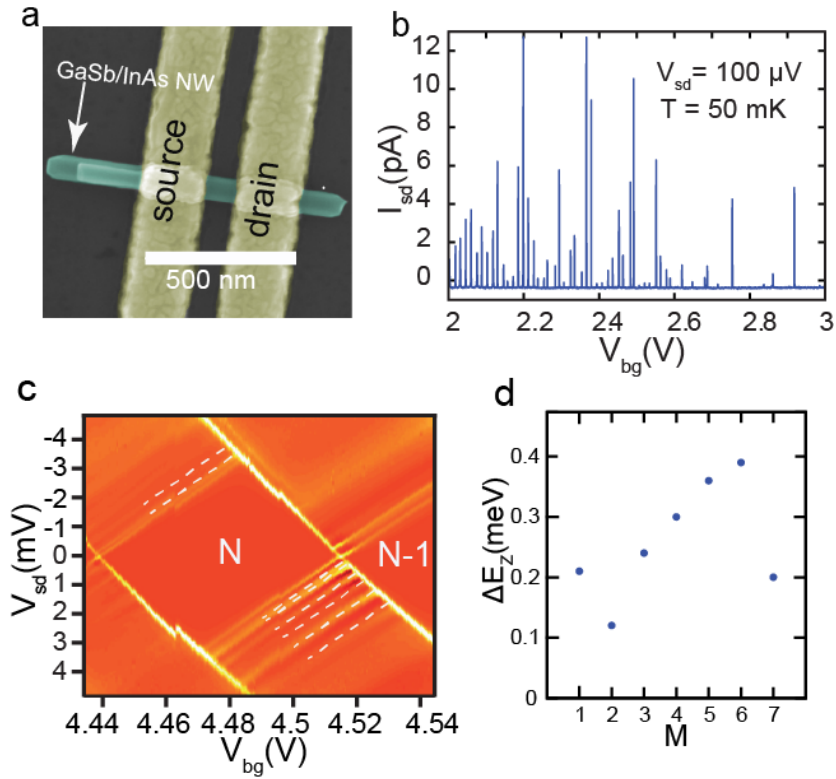

Figure S3: (a) SEM image of device $B$. (b) Coulomb blockade oscillations as a function of the back-gate voltage at source–drain bias voltage of $V_{sd}= 100\ \mu$V measured at T= 50 mK. The Coulomb peak spacing decreases and the current increases with decreasing gate voltage, which supports the interpretation that a hole quantum dot has formed between the source and drain contacts. (c) High-resolution charge stability diagram of the $N$th Coulomb diamond in Fig 3(a) in the main paper. The high conductance lines running parallel to the diamonds borderlines, indicated with dashed lines, correspond to excited states. (d) Single particle energy level-spacings ($\Delta$E) extracted from the parallel lines on the lower-right side of the $N$th diamond (giving an average spacing of 0.26 meV).

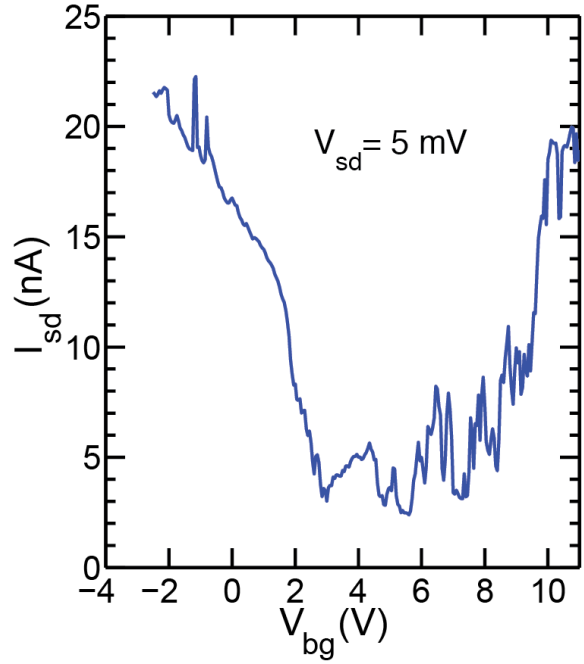

Figure S4: *I-V*$_{bg}$ measured for another GaSb/InAs core-shell nanowire quantum dot (Device *C*). The base temperature is 12 mK. The device shows ambipolar conduction properties with dominating hole and electron conduction for negative and positive gate voltages, respectively. Holes and electrons likely coexist in a large gate voltage span of 2 to 8 V.

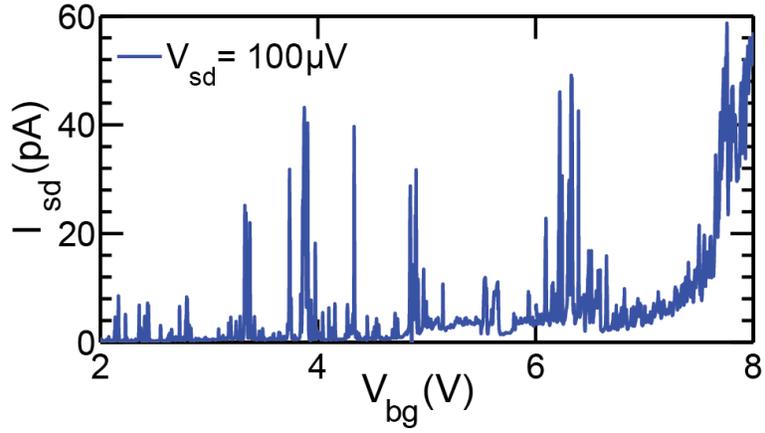

Figure S5: Coulomb blockade oscillations in conductance as a function of the back-gate voltage at source–drain bias voltage of $V_{sd} = 100 \ \mu$V for Device *C* showing no clear cross-over from hole- to electron-dominated transport. Spatially separated holes and electrons may coexist for gate voltages in this range.

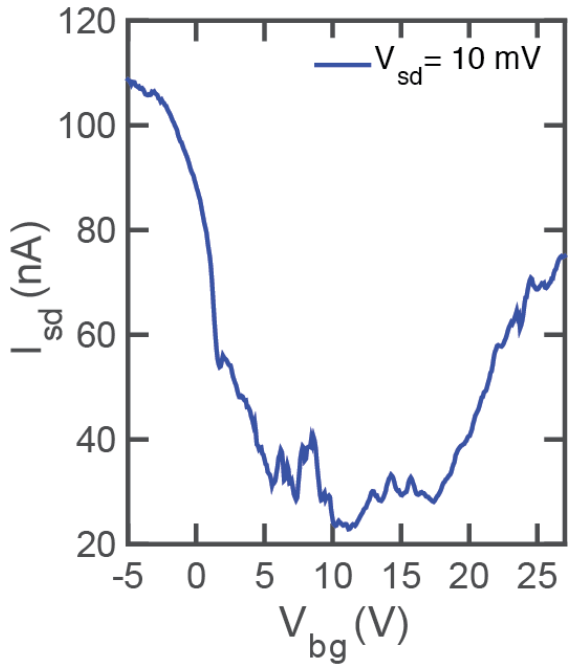

Figure S6: *I-$V_{bg}$* measured for another GaSb/InAs core/shell nanowire quantum dot (Device *D*). The device shows ambipolar conduction properties with a minimum conduction point around 10 V. The conductance remains significant also at this minimum, indicating that holes and electron coexist over some range of gate voltages.

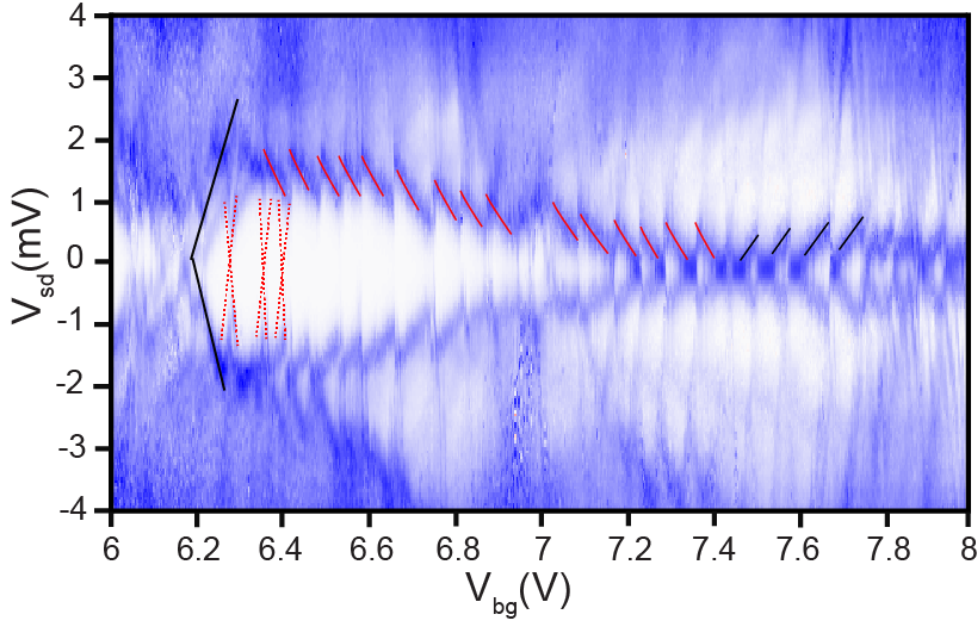

Figure S7: $dI/dV_{sd}$ versus $V_{sd}$ and $V_{bg}$, measured for Device $D$ close to the transition region from holes to electrons. The measurements show two patterns of diamonds with different slopes super-imposed on each other. Based on the analysis in Fig. 4, we interpret lines with small slope to be associated with hole transport (marked with solid lines) and lines with large slope to be associated with electrons (dashed lines). Each time an electron is added, the resonance for adding (removing) a hole jumps to lower (higher) bias voltages, giving rise to discontinuities in the resonance lines, marked with black (red) solid lines.

**Magneto-transport measurements:** For Device $B$, we have also studied how the Zeeman-energy increases with increasing magnetic field. A perpendicular magnetic field lifts the twofold spin-degeneracy of the ground states with an odd number of holes in the quantum dot, and hence it splits into a spin-up and a spin-down resonance, indicated by blue arrows in Fig. S7(a-d). Two spin states are separated in energy by the Zeeman energy, $E_Z = |g*|\mu_B B$, where $g*$ is the $g$-factor and $\mu_B = 5.8 \times 10^{-5}$ eV T$^{-1}$ is the Bohr magneton. However, as shown in Fig. S8(e) the Zeeman splitting does not increase linearly with the applied magnetic field. It saturates at higher applied magnetic fields ($B > 2$T) because of an anti-crossing due to spin-orbit coupling as also shown in Fig. 4(b) and Fig. S10(a).

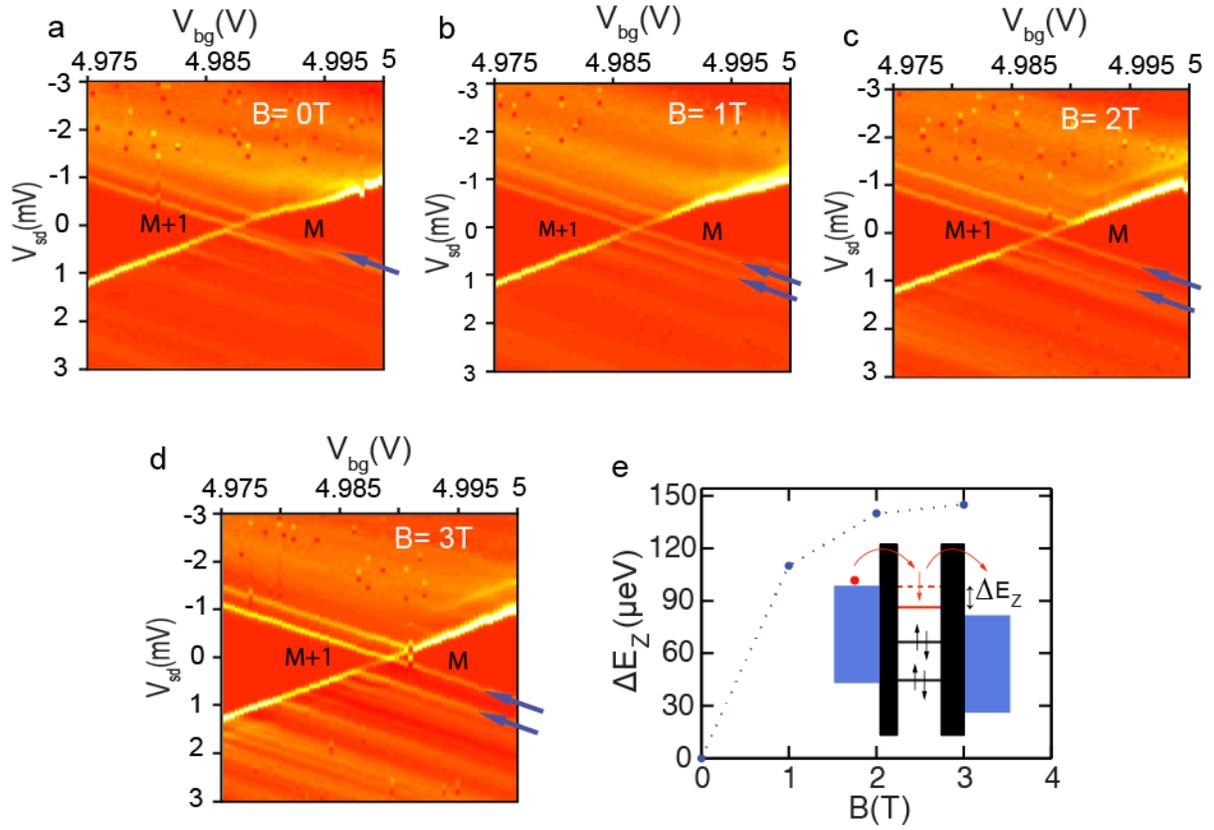

Figure S8: (a) $dI/dV_{sd}$, versus $V_{sd}$ and $V_{bg}$, measured at zero magnetic field. (b-d) Charge stability diagram measured at magnetic fields $B$=1, 2, and 3 T. The Zeeman splitting does not increase linearly with the applied magnetic field because of an anti-crossing with an excited quantum dot level, indicating a presence of an interaction in the nanowire quantum dot. (e) The Zeeman splitting, $\Delta E_Z$, versus magnetic field, $B$.

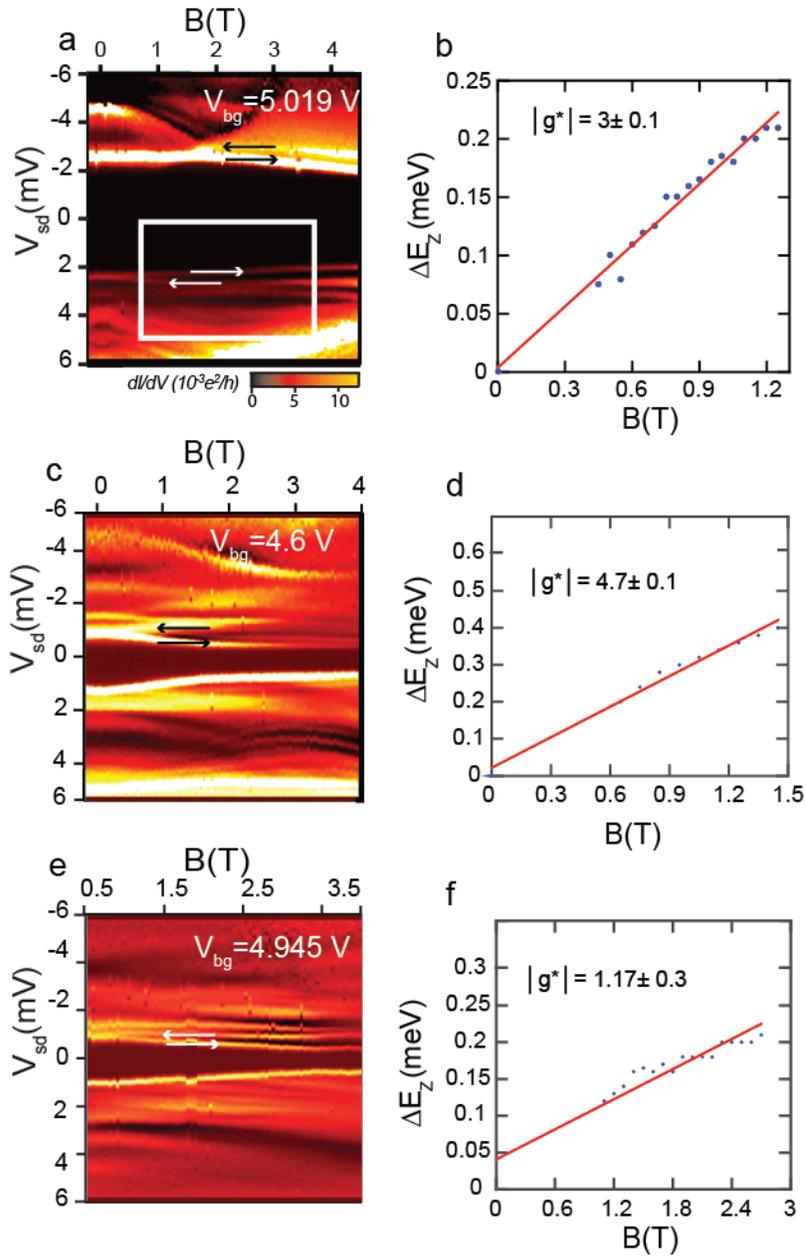

Figure S9: (a,c and e) $dI_{sd}/dV_{sd}$ as a function of $V_{sd}$ and $B$ at $V_{bg} = 5.019$ V, 4.6 V and 4.945V . (b, d and f) Zeeman splitting extracted from the ground state splitting in panel (a) versus magnetic field. A linear fit to the Zeeman splitting extracted from the ground state splitting in panel (a, c and e) yield values of |$g$*|= 3±0.1, 4.7±0.1 and 1.17±0.3, respectively.

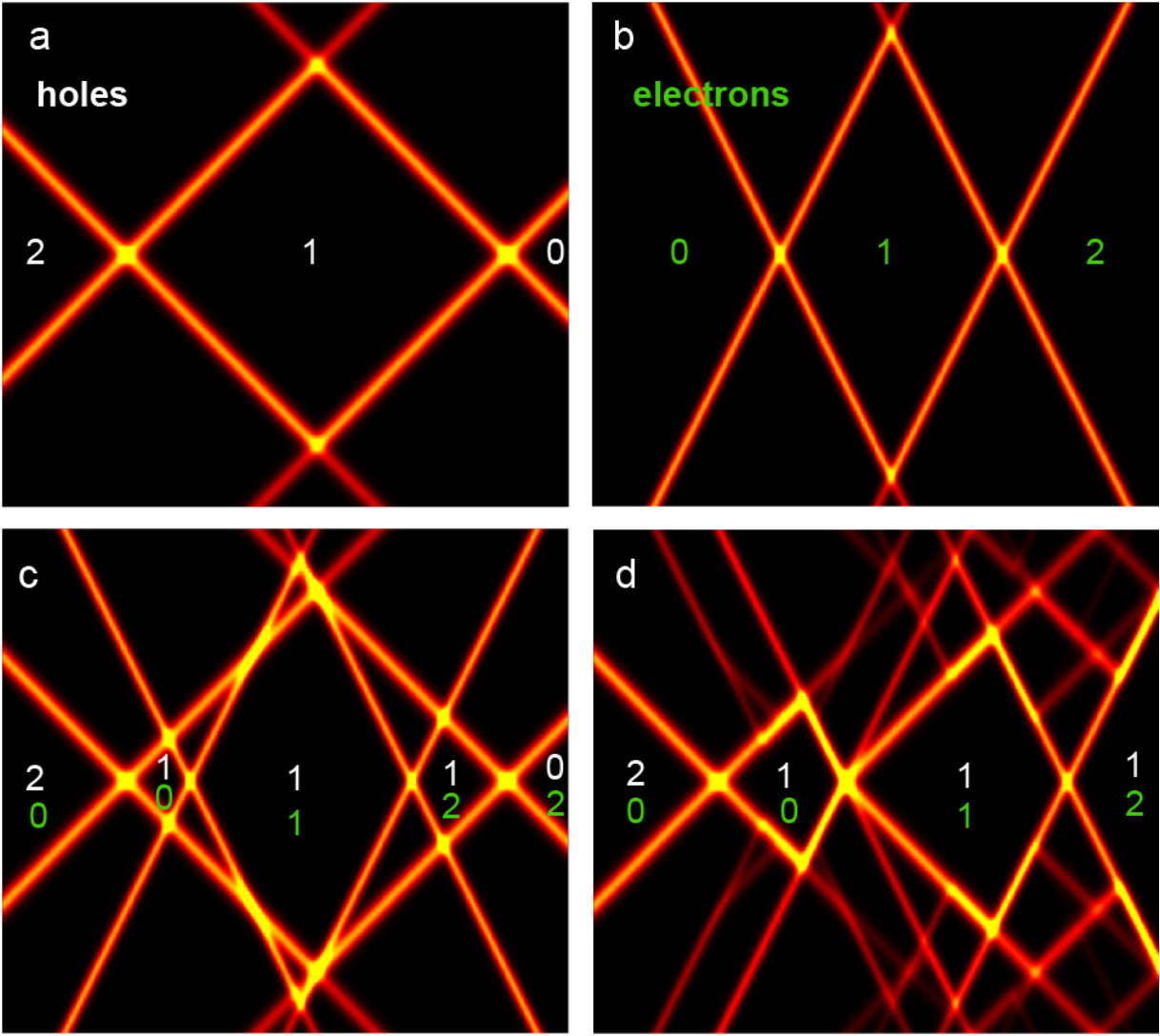

Figure S10: Simulated stability diagram of (a) a hole quantum dot, (b) an electron quantum dot, (c) a parallel electron and hole quantum without considering the interaction between electron and hole states, and (d) a parallel electron and hole quantum dot with electron-hole interaction (same as Fig. 3(d) in the main paper). The numbers show the electron (green) and hole (white) occupation numbers, with the electron and hole numbers defined as $n_e=0$ and $n_h=2$, respectively, at the far left.

**Modeling details:** We use the simplest possible model which can capture transport through the interacting electron and hole double quantum dot. Only a single spin-degenerate level is included in each dot, which has energy $\varepsilon_h$ ($\varepsilon_e$) and occupation $n_h$ ($n_e$) for the hole (electron) dot. The double dot system is then described by the Hamiltonian $H=\varepsilon_h n_h+U_h n_h(n_h-1)+\varepsilon_e n_e+U_e n_e(n_e-1)-\Delta_{e-h} n_e n_h$, where $U_h$ and $U_e$ are the strengths of the repulsive local Coulomb interactions on the hole and electron dots, respectively, and $\Delta_{e\text{-}h}$ is the strength of the attractive non-local Coulomb interaction between electrons and holes. Both the electron and hole dots are tunnel coupled to metallic source and drain contacts and good agreement with the experiment is found when the hole and electron dots have equal tunnel couplings, and when each dot has equal tunnel couplings to both the source and drain. The nonequilibrium occupations of the double quantum dot and the total current are calculated using rate equations based on lowest order perturbation theory in the tunnel couplings (sequential tunneling approximation). In the results presented here we have used $U_h=75T$, $U_e=62.5T$, and $\Delta_{e\text{-}h}=25T$, where $T$ is the temperature of the source and drain contacts. Furthermore, the gate coupling of the hole quantum dot is set to half that of the electron dot.